\documentclass{ws-ijmpd}

\begin{document}

\markboth{Julien Malzac}
{The X-ray corona and jet of Cygnus X-1}

%
\catchline{}{}{}{}{}
%

\title{THE X-RAY CORONA AND JET OF CYGNUS X-1  }

\author{JULIEN MALZAC}
\author{RENAUD BELMONT}

\address{Centre d'Etude Spatiale des Rayonnements, CNRS (UMR5187),  Universit\'e de Toulouse (UPS), 9 Avenue du Colonel Roche, BP44346, 31028 Toulouse Cedex 4, France\\
malzac@cesr.fr}

\maketitle

\begin{history}
\received{Day Month Year}
\revised{Day Month Year}
\comby{Managing Editor}
\end{history}

\begin{abstract}
Evidence is presented indicating that in the hard state of Cygnus X-1, the coronal magnetic field might be below equipartition with radiation (suggesting that the corona is not powered by magnetic field dissipation) and that the ion temperature in the corona is significantly lower than what predicted by ADAF like models.  It is also shown that the current estimates of the jet power set interesting contraints on the jet velocity (which is at least mildly relativistic), the accretion efficiency (which is large in both spectral states), and the nature of the X-ray emitting region (which is unlikely to be the jet).
\end{abstract}


\section{Introduction}\label{sec:intro}

 Black hole binaries are observed in two main X-ray spectral states, namely the Hard State (HS) and the Soft State (SS), see [1]. The hard X-ray emission in both spectral states is well represented by Comptonisation by an hybrid thermal/non-thermal electron distribution. In the HS the temperature and optical depth of the thermal electrons are higher, and the slope of the non-thermal tail seem steeper than in the SS.  Consequently, the hard X-ray emission is dominated by thermal Comptonisation in the HS and by non-thermal Comptonisation in the SS.  The HS is known to be associated with the presence of a compact radio jet which   observed in the SS.  Here we focus  on the prototypical black hole source Cygnus X-1.
In section~\ref{sec:model}, we present a relatively simple coupled kinetic-radiation model that  allows us to understand the origin of the very different spectral shapes observed in the two spectral states as well as  the spectral evolution during state transitions (see e.g. [2]). A thorough investigation of the model and its discussion in the context of the observations  can be found  in [3]; the present paper summarises our main results.  
Then in section~\ref{sec:jet}, we summarise the arguments developed in [4] showing that the present estimates of the jet power of Cygnus X-1 imply that the jet has a relativistic velocity, that the accretion proceeds efficiently in the HS and that the X-ray emission is unlikely to be produced in the jet.

\section{Magnetic field and ion temperature in the corona}\label{sec:model}
\subsection{Model}
The code of [5] solves the kinetic equations for photons, electrons and positrons in the one-zone approximation. It accounts for Compton scattering (using the full Klein-Nishima cross section), $e^{+}$-$e^{-}$ pair production and annihilation, Coulomb interactions (electron-electron and electron-proton), synchrotron emission and absorption and $e$-$p$ bremsstrahlung. Radiative transfer is dealt using a usual escape probability formalism (Poutanen, in these proceedings, for the presentation of a very similar code).

We model the Comptonising region as a sphere with radius $R$  of fully ionised proton-electron magnetised plasma in steady state. The Thomson optical depth of the sphere is $\tau_{\rm T}=\tau_{\rm i}+\tau_{\rm s}$, where $\tau_{\rm i}=n_i\sigma_{\rm T}R$ is the optical depth of ionisation electrons (associated with protons of density $n_i$) and $\tau_s=2n_{\rm e^{+}}\sigma_{\rm T}R$ is the optical depth of electrons and positrons due to pair production ($n_{\rm e^{+}}$ is the positron number density). $\sigma_{\rm T}$ is the Thomson cross section.
 The radiated power is quantified through the usual compactness parameter  $l=\frac{L\sigma_{\rm T}}{R m_{\rm e} c^3}$,
where $L$ is the luminosity of the Comptonising cloud,  $m_{\rm e}$ the electron rest mass and $c$ the speed of light. 

We consider three possible channels for the energy injection in our coupled electron-photon system: { (i) Non-thermal electron acceleration. We model the acceleration process by assuming electrons are continuously re-injected with a power-law distribution of index $\Gamma_{\rm inj}$. The power provided to the plasma is parametrised  by the non-thermal compactness $l_{\rm nth}={L_{\rm nth}\sigma_{\rm T}}/{(R m_{\rm e} c^3)}=[\langle\gamma\rangle_{\rm i}-\langle\gamma\rangle]\dot{n} \sigma_{\rm T}/(Rc)$,
where $L_{\rm nth}$ is the power provided to electrons through the acceleration process, $\dot{n}$ is the total number of injected/accelerated particles per unit time,  $\langle\gamma\rangle$  is the average Lorentz factor of electrons in the hot plasma,  and $\langle\gamma\rangle_{\rm i}$ is the average Lorentz factor at which they are accelerated/injected. 
(ii) Coulomb heating. Electrons are supposed to interact by Coulomb collisions with a distribution of thermal ions. The ions are supposed to be heated by some unspecified process.  If the ions have a larger temperature than the electrons (as  in two-temperature accretion flows), Coulomb collisions will heat the electrons.  For  $L_{c}$  the power transferred from ions to the leptons through Coulomb collisions, we define the electron heating compactness $l_{\rm c}=\frac{L_{\rm c}\sigma_{\rm T}}{R m_{\rm e} c^3}$.
In our model, $l_c$ is the parameter, and  the temperature of the ions is determined accordingly so that the electron heating is $L_{\rm c}$. For $l_{\rm c}=0$ the two populations have the same temperature. 
(iii)  Soft photons injection.  External soft radiation coming from the geometrically thin accretion disc, may enter the corona with a compactness  $l_{\rm s}=\frac{L_{\rm s}\sigma_{\rm T}}{R m_{\rm e} c^3}$, where $L_{\rm s}$ is the soft photon luminosity entering the corona. } 
Since, in this model, all the injected power ends up into radiation we have: $l=l_{\rm nth}+l_{\rm c}+l_{\rm s}$ in steady state. In addtition the magnetic field $B$ is parametrized through the usual magnetic compactness: $l_{B}=\frac{\sigma_{\rm T}}{m_{\rm e} c^2}R \frac{B^2}{8\pi}$.

{ Once the injection parameters are specified, the code computes the steady state particle distribution and photon spectrum escaping from the corona. The injected non-thermal particles cool down though Compton and synchrotron radiation and thermalise through $e$-$e$ Coulomb collisions and synchrotron self-absorption. The result at steady state is a Maxwellian distribution at low energies and, at higher energies a power law tail made of the cooling particles.}

\begin{figure}[t]
\centering
\includegraphics[width=8cm]{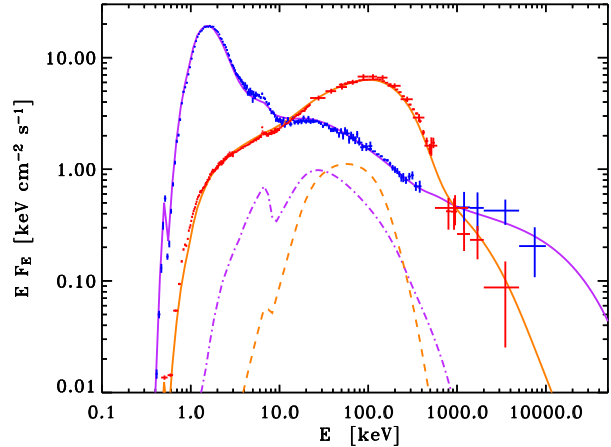}
\caption{A comparison of the average CGRO spectra for the SS and  HS of Cygnus X-1 (crosses, data from [8]), with models (solid lines)  involving only injection of non-thermal particles as sole heating mechanism. At low energy the CGRO data are complemented by BeppoSAX. Reflection components were added to both spectra and are shown by the thin dot-dashed and dashed curves for the SS model and the HS model respectively. \label{fig:lcygdat}}
\end{figure}

\subsection{Results}

We find that a pure non-thermal injection model (i.e. assuming $l_{\rm c}=0$)  provides a good description of the high energy spectra of Cygnus X-1 in both spectral states (see Fig~\ref{fig:lcygdat}).  According to our  models, the non-thermal compactness of the corona is comparable ($l_{\rm nth}\simeq5$)  in both spectral states. As expected most of the differences between HS and SS are due to a change in the soft photon compactnes $l_{\rm s}$ that we assumed to be 0 and about 3$l_{\rm nth}$ in the HS and SS respectively.  { In the HS the synchrotron and $e$-$e$ Coulomb boilers redistribute the energy of the non-thermal particles to from and keep a quasi-thermal electron distribution at a relatively high temperature, so that most of the luminosity is released through quasi thermal Comptonisation. In the SS,  the soft photon flux from the accretion disc becomes very strong and cools down the electrons, reducing the thermal Compton emissivity. In the SS the hard X-ray emission is then dominated by inverse Compton on the non-thermal particles forming the power law observed up to a few MeV  in this state.}

We find the magnetic field must be relatively low in the HS (while it is not very well constrained in the SS). In fact if we compare the ratio of the magnetic energy density to the total radiation energy density (related to $l$) in both models we find the  $U_{B}/U_{\rm R}\simeq3$ in the SS while in the HS,  $U_{B}/U_{\rm R}\simeq0.3$.
In any case, the magnetic field inferred from our model in the HS is probably an upper limit on the actual magnetic field in the source. The presence of external soft photons (neglected in this fit) would imply a lower $B$ to keep the coronal temperature high.
The fact that this maximum magnetic field  appears significantly below equipartition with radiation suggests that the emission of the corona is not powered by the magnetic field, as assumed in most accretion disc corona models (e.g. [6]; [7]). { We note that this constraint is a general conclusion of our modelling. It does not depend on the details of the model of the power supply to the electrons (Coulomb or non-thermal) acceleration. It comes form the presence of a non-thermal high energy tail above the thermal Comptonisation cut-off in the HS of Cygnus X-1 [8]. This power-law tail implies the presence of a known number of non-thermal high energy Comptonising electrons which also produce self-absorbed synchrotron emission in the IR-optical bands. This soft radiation is Comptonised and tend to cool down the corona. If the synchrotron luminosity is too strong then it is impossible to sustain the relatively high ($\sim$ 100 keV) temperature of the Maxwellian electrons, and therefore the magnetic field must be low.  The constraint would  however be relaxed if the non-thermal excess in the photon spectrum was produced in a different region than the bulk of  the thermal Comptonisation luminosity. We note that very similar results were obtained independently by [9]. }

Alternatively, models with heating by hot protons (i.e. $l_{\rm c}>0$) also provide a very good description of the spectra of Cygnus X-1. However even in these models some level of non-thermal acceleration is required in order to reproduce the non-thermal MeV tails. In our 'best fit' models about 25 \% of the heating power is provided in the form of non-thermal acceleration in the HS while this fraction rises to about 2/3 in the  SS.  We infer ion temperature of about 50 MeV in the SS versus  only 1.3 MeV in  the HS.  We note that in the HS the proton temperature appears significantly lower than what predicted by standard 2-temperature accretion flow solutions (the temperature of the hot protons in typical ADAF models is in the range 10--100 MeV). {  This difference comes from the large Thomson depth ($\tau_T>1$) that is required to fit the spectrum, while ADAF models always predict  $\tau_{\rm T}<<1$. As a consequence the Coulomb coupling between ions and electrons is much more efficient in our calculation. A large temperature of the protons would produce a luminosity that is larger than observed.  The proton temperature could even be much lower than 2 MeV, if, as it is likely, there is an additional electron heating mechanism beside $e$-$p$ Coulomb. Independently of the details of the model, this indicates that the temperature of ions and electrons are comparable in Cygnus X-1, differing at most by a factor $\sim 20$.} 

\section{Constraints from the jet power}\label{sec:jet}
Deep radio observations of the field of Cygnus X-1 resulted in the discovery of a shell-like nebula
which is aligned with the resolved radio jet ([10]; [11]). This large-scale (5 pc in diameter) structure appears to be inflated by the inner radio jet. In order to sustain the observed emission of the shell, the jet of Cygnus X-1 has to carry a kinetic power that is comparable to the bolometric X-ray luminosity  of the binary system [11]. 
This estimate was refined in [12] and it was found that the total kinetic power of the double sided jet  is $L_{\rm J}$=(0.9--3) $\times 10^{37}$~erg~s$^{-1}$.
If we adopt $L_{h}$=2 $\times$ 10$^{37}$~erg~s$^{-1}$ as the typical X-ray luminosity in the HS then $j=L_{\rm J}/L_{\rm h}$ is in the range 0.45--1.5.
In [4], we  showed that this estimate of the jet power sets some constraints on the physics of the accretion and ejection in Cygnus X-1 that we sumarise below.

\subsection{The accretion efficiency and jet velocity}
In the SS there is overwhelming evidence that accretion proceeds through a geometrically thin optically thick accretion disc. Therefore the accretion is radiatively efficient. Assuming that there is no jet in the SS,  the efficiency can be defined as:
\begin{equation}
\eta_{\rm s}=L_{\rm s}/\dot{M}_{\rm s}c^2,
\label{eq:etas}
\end{equation}
where $L_{\rm s}$ is the source  luminosity in the SS, $\dot{M}_{\rm s}$ the mass accretion rate.
According to the theory of general relativity $\eta_{\rm s}$ is in the range 0.06--0.4 depending on the spin of the black hole.  
In the HS,  the accretion probably does not proceed through a standard thin disc and the efficiency is essentially unknown. 
Depending on the nature of the accretion flow, it could be close to that of the SS or much smaller (as e.g. in an advection dominated accretion flow). If we take into account the presence of the energetically important jet,  the efficiency in the HS can be formaly writen as:
\begin{equation}
\eta_{\rm h}=\frac{L_{\rm J}+L_{\rm h}}{(1-f_{\rm j})\dot{M}_{\rm h}c^2},
\label{eq:etah}
\end{equation}
where  $f_{\rm j}$ represents the fraction of the accreting material which is ejected and cannot be used to release energy.
We know from the observations that  $L_{\rm j}\simeq L_{\rm h}$ and that $L_{\rm s}/L_{\rm h}<4$.  Moreover in Cygnus X-1 the spectral transition is believed to be triggered by an increase in the accretion rate and therefore $\dot{M}_{\rm s}>\dot{M}_{\rm h}$. Combining this with equations~(\ref{eq:etas}) and (\ref{eq:etah}), it follows that $\eta_{\rm h}>\eta_{\rm s}/2$. This shows that accretion is quite efficient in the HS and cannot be strongly advection dominated. 

The jet kinetic power can be written as:
\begin{equation}
L_{\rm J}=f_{\rm j}\dot{M}_{\rm h}(\gamma_{\infty}-1)c^2,
\end{equation}
combining this with equations~(\ref{eq:etah}) and~(\ref{eq:etas}) we get an estimate of the terminal bulk Lorentz factor of the jet:
\begin{equation}
\gamma_{\infty}=1+\frac{j\eta_{\rm s}}{\lambda-(1+j)\frac{\eta_{\rm s}}{\eta_{\rm h}}}
\end{equation}
where $j=L_{\rm j}/L_{\rm h}\simeq 1$ and $\lambda=\frac{L_{\rm s}}{L_{\rm h}}\frac{\dot{M}_{\rm h}}{\dot{M}_{\rm s}}$.

{ The observations constrain the ratio $\frac{L_{\rm s}}{L_{\rm h}}\simeq 3-4$.  Cygnus X-1 and other sources   are observed most of the time either in the HS or SS  and very rarely in intermediate states, which appear very unstable. If $\dot{M}_{\rm h}/\dot{M}_{\rm s}$ is too small, it is  difficult to understand why the source sytematically avoids the 'intermediate' range of mass accretion rates.  For this reason we expect  $\dot{M}_{\rm h}/\dot{M}_{\rm s}$ to be close to unity and $\lambda\simeq3$. }

For typical parameters  (e.g. $\lambda\simeq3 $ and $\eta_{\rm s}\simeq\eta_{\rm h}\simeq0.1$), this gives estimates of the terminal jet velocity of 0.4$c$ (see [13] for a full investigation of the parameter space). Such mildly relativistic velocities are in agreement with independent estimates based on radio observations (see [13]; [14]).
An absolute lower limit on the jet velocity $\beta_{\infty}=v_{\infty}/c>0.1$ is obtained in the extreme case $\eta_{\rm h}=1$, $\eta_{\rm s}=0.06$, $\lambda=4$ and $j=0.45$. The jet velocity is therefore at least mildly relativistic.

\subsection{Does the X-ray emission originate in the jet ?} 

Several authors have suggested that the X-ray  may actually be produced by the jet or its base (e.g. [15], [16], [17]). The jet energetics shows however that this is unlikely.
Indeed, mass flux is conserved along the jet, and therefore the rate at which mass is ejected can be written as:
\begin{equation}
\dot{M}_{\rm J}=\frac{L_{\rm J}}{(\gamma_{\infty}-1)c^2}=\pi r(z)^2 n(z) \beta(z) \gamma(z) m_{\rm p} c < 6.6 \times 10^{18} \quad {\rm g \quad s}^{-1},
\label{eq:mdotj}
\end{equation}
where $r$, $n$, $\beta$ and $\gamma$ represent the jet section, density, bulk velocity and Lorentz factor at a given height $z$ above the black hole. 
The upper limit of  6.6 $\times$ 10$^{18}$  g s$^{-1}$ comes from the constraints  $\beta_{\infty}>0.1$ and $L_{\rm J}<3 \times 10^{37}$ erg s$^{-1}$.

{  The X-ray emission in the HS  is unlikely to be dominated  by the emission of a \emph{non-thermal} electron population\footnote{Here we do not mean, as before, non-thermal injection but that the final population is not thermalized.}.  None of the current models based on this idea is able to provide a good description of the Cygnus X-1 data. The original non-thermal synchrotron jet model of [18] reproduces the spectrum of XTEJ1118+480.  However, the fit of the high energy spectrum of Cygnus X-1 is not as good as that obtained with Comptonisation models. The synchrotron model of the HS of Cygnus X-1 (see fig. 3a in [19]) when matched to the 100 keV flux overestimates the 1 MeV flux (see Fig.~ 1) by a factor of 8. It seems that only the most recent version of the model [17] including an additional \emph{thermal}  SSC component from the base of the jet can fit the spectrum of Cygnus X-1 satisfactorily.  Similarly,  the non-thermal IC jet model of [15] would require a jet power of~$\sim$ 5 10$^{38}$ erg s$^{-1}$ which is immediately ruled out by the estimates of [12]. 

In contrast, thermal (or quasi-thermal) Comptonisation models have been shown to provide a very good description of the high energy  spectra of all observed black hole (and neutron star) binaries in the HS [1].} Those spectral fits with thermal Comptonisation models of  the HS high energy spectra of Cygnus X-1 require a Thomson optical depth $\tau_{\rm T}$ in the range 1--3. 
If we assume that this Comptonised emission is produced somewhere in the jet  at some height $z_0$. This gives the additional constraint:
\begin{equation}
\tau_T\simeq n(z_0)\sigma_T r(z_0) >1
\label{eq:taujet}
\end{equation}
where $\sigma_{\rm T}$ is the Thomson cross section. 
Combining this with the constraints from equation~(\ref{eq:mdotj}), we find that for any reasonable jet section (i.e. $> 10 R_G$) the velocity of the X-ray emitting region 
must be non relativistic, $\beta(z_0)<0.1$.  In fact this upper limit is very conservative  and for 
reasonable parameters we actually expect the velocity to be much lower than 0.1c (see [4]). The part of the jet producing the X-rays should be very slow and  most of the jet bulk velocity would have to be acquired at larger distances from the black hole. This is a possibility specially if the jet is magnetically accelerated (see Narayan et al.  these proceedings).  However, most X-ray jet models require or assume both $\tau_{T}>1$ and initial velocity of the X-ray emitting region that is larger than 0.1 $c$. As long as the estimates of [12] are correct, these models can be ruled out. This conclusion however does not apply to the jet model of [17].  Indeed, in this model the X-ray emission is produced 
through synchrotron self-Compton emission of very energetic electrons of temperature of a few MeV and optical depth $\tau_{\rm T}\sim10^{-3}-10^{-2}$. 
The low Thomson depth makes it  energetically possible to have a mildly relativistic initial jet velocity. We note however that such a combination of small size, very low optical depth and large temperature is physically impossible.  Indeed in  Cygnus X-1, the large luminosity and small emitting region make the compactness larger than unity and electron positron production can be significant. For a compactness of order of a few (like in Cyg X-1), the optical depth must be at least 10 times larger than in this jet model (see discussion in [4]). { The parameters of the model of [17] therefore appear inconsistent with the constraints from pair equilibrium.} 

Finally we note that although the jet is unlikely to contribute significantly to the hard X-ray emission of Cygnus X-1, it may nevertheless be at the origin of the gamma-ray emission that was detected by the MAGIC telescope ([20]; [21]; [22]).

\section{Conclusions}
 In both spectral states of black hole binaries the 
 coronal emission  can be powered by a similar non-thermal acceleration mechanism. In the HS the synchrotron and $e$-$e$ Coulomb boilers redistribute the energy of the non-thermal particles to form and keep a quasi-thermal electron distribution at a relatively high temperature, so that most of the luminosity is released through quasi-thermal Comptonisation. In the SS,  the soft photon flux from the accretion disc becomes very strong and cools down the electrons, reducing the thermal Compton emissivity. This change in the soft photon flux could be caused either because the inner radius of the truncated disc moves inward into the central hot accretion flow, or, in the framework of accretion disc corona models, because the disc temperature increases dramatically. Then most of the emission is produced by disc photons up-scattered by the non-thermal cooling electrons.  
Our comparison of simulations with the high energy spectra of Cygnus X-1 in the HS allowed us to set upper limits on the magnetic field and the proton temperature.  Our results indicate that, { as long as the non-thermal MeV tail is produced in the same region as the bulk of the Comptonised emission}, the  magnetic field  in the HS  is below equipatition with radiation (unlike what is assumed in most accretion disc corona models). The proton temperature is found to be significantly lower than predicted by standard 2-temperature accretion flow models ($kT_{\rm i}<$2 MeV). We also note that such accretion flows are usually radiatively inneficient while the jet energetics suggests efficient accretion in the HS. The present estimates of the jet power also suggest that the jet is Thomson thin with a velocity which is {at least} mildly relativistic, which does not make it a favoured location for the production of the observed X-ray emission.  

\section*{Acknowledgments}
JM is grateful to the Institute of Astronomy in Cambridge (UK) for hospitality. This research was funded by CNRS and ANR. 
The authors thank the anonymous referee for providing them with useful comments that lead to significant improvements.

\end{document}